\documentclass[journal]{IEEEtran}
\usepackage{acronym}
\usepackage{amsmath,amsfonts}
\usepackage{tikz}
\newcommand{\Cross}{\mathbin{\tikz [x=1.4ex,y=1.4ex,line width=.2ex] \draw (0,0) -- (1,1) (0,1) -- (1,0);}}%
\usepackage{algorithmic}
\usepackage[flushleft]{threeparttable}
\usepackage{algorithm}
\usepackage{array}
\usepackage{booktabs}
\usepackage[caption=false,font=normalsize,labelfont=sf,textfont=sf]{subfig}
\usepackage{textcomp}
\usepackage{stfloats}
\usepackage{url}
\usepackage{verbatim}
\usepackage{graphicx}
\usepackage{cite}
\IEEEoverridecommandlockouts
\usepackage{cite}
\usepackage{amsmath,amssymb,amsfonts,mathtools}
\usepackage{algorithmic}
\usepackage{graphicx}
\usepackage{textcomp}
\usepackage{multirow}
\usepackage{multicol}
\usepackage{makecell}
\usepackage{enumitem} 
\usepackage{authblk} 
\usepackage[bookmarks=false]{hyperref}
\usepackage{academicons}
\definecolor{lime}{HTML}{A6CE39}
\DeclareRobustCommand{\orcidicon}{%
	\begin{tikzpicture}
	\draw[lime, fill=lime] (0,0) 
	circle [radius=0.16] 
	node[white] {{\fontfamily{qag}\selectfont \tiny ID}};
	\draw[white, fill=white] (-0.0625,0.095) 
	circle [radius=0.007];
	\end{tikzpicture}
	\hspace{-2mm}
}
\foreach \x in {A, ..., Z}{%
	\expandafter\xdef\csname orcid\x\endcsname{\noexpand\href{https://orcid.org/\csname orcidauthor\x\endcsname}{\noexpand\orcidicon}}
}

\begin{document}

\title{Experiment-based Models for Air Time and Current Consumption of LoRaWAN LR-FHSS}

\author{Muhammad Asad Ullah\textsuperscript{1}, Konstantin Mikhaylov\textsuperscript{2}, and Hirley Alves\textsuperscript{2}}
\affil{\textsuperscript{1}VTT Technical Research Centre of Finland Ltd., Espoo, Finland\\ \textsuperscript{2}Centre for Wireless Communications, University of Oulu, Oulu, Finland\\
Email: \textsuperscript{1}asad.ullah@vtt.fi, \textsuperscript{2}firstname.lastname@oulu.fi
\thanks{This work has been submitted to the 
IEEE Internet of Things Journal for possible publication. Copyright to IEEE may be transferred without notice.}}

\maketitle

\vspace{-30pt}
\begin{abstract}
Long Range - Frequency Hopping Spread Spectrum (LR-FHSS) is an emerging and promising technology recently introduced into the LoRaWAN protocol specification for both terrestrial and non-terrestrial networks, notably satellites.
The higher capacity, long-range and robustness to Doppler effect make LR-FHSS a primary candidate for direct-to-satellite (DtS) connectivity for enabling Internet-of-things (IoT) in remote areas. The LR-FHSS devices envisioned for DtS IoT will be primarily battery-powered. Therefore, it is crucial to investigate the current consumption characteristics and Time-on-Air (ToA) of LR-FHSS technology. However, to our knowledge, no prior research has presented the accurate ToA and current consumption models for this newly introduced scheme. This paper addresses this shortcoming through extensive field measurements and the development of analytical models. \textcolor{black}{ Specifically, we have measured the current consumption and  ToA for variable transmit power, message payload, and two new LR-FHSS-based Data Rates (DR8 and DR9).}  We also develop current consumption and ToA analytical models demonstrating a strong correlation with the measurement results exhibiting a relative error of less than 0.3\%. Thus, it confirms the validity of our models. Conversely, the existing analytical models exhibit a higher relative error rate of -9.2 to 3.4\% compared to our measurement results. The presented in this paper results can be further used for simulators or in analytical studies to accurately model the on-air time and energy consumption of LR-FHSS devices. 
\end{abstract}

\vspace{-10pt}
\section{Introduction}
\textcolor{black}{The LoRaWAN direct-to-satellite (DtS) communication is emerging as a potential connectivity solution for IoT applications, especially for the ones deployed in remote areas~\cite{TOAES,TOAES1}. As a recent advancement to the LoRaWAN protocol specification, Long Range - Frequency Hopping Spread Spectrum (LR-FHSS) has been designed and introduced to support both terrestrial and non-terrestrial IoT networks to offer low-power connectivity. 
Along with the long coverage, LR-FHSS offers robustness to interference through lower code rate, intra-packet frequency hopping, and header diversity~\cite{1}. Due to these characteristics and strong robustness to the Doppler effect, LR-FHSS stands out as one of the prominent low-power wide-area network (LPWAN) technologies to enable direct connectivity between machine devices and the Low Earth Orbit (LEO) satellites.} In~\cite{Datasheet1}, an initial insight into the LR-FHSS network performance and theoretical capacity has been provided. Specifically, in European Telecommunications Standards Institute (ETSI) regions, a single gateway operating on 125~kHz bandwidth can receive up to 1~million daily packets while maintaining a 10\% error rate. Conversely, according to the Federal Communications Commission (FCC) region with 1.5~MHz channel bandwidth, the capacity increases to 11~million daily packets. The LR-FHSS technology aligns with the connectivity needs of DtS IoT by offering a combination of high capacity, resistance to Doppler effects, and robustness against interference~\cite{12}. The Long Range Wide Area Network (LoRaWAN) protocol, the LR-FHSS, and conventional LoRa modulations can co-exist without modifying the existing network architecture. To leverage this incorporation, Adaptive Data Rate (ADR) commands from the network server are sufficient to switch between the modulations~\cite{2}.  However, LR-FHSS only supports uplink communication, i.e., from the device to a gateway. For the successful demodulation of an uplink LR-FHSS packet, a LoRaWAN V2 gateway \textcolor{black}{with digital signal processing capabilities} is required. The network can employ either LoRa or Frequency-shift keying (FSK) modulations for downlink communication~\cite{USerManual}.

\subsection{Relevant works}
 The earlier studies discuss and examine LR-FHSS DtS performance, mainly focusing on network scalability. Specifically, study~\cite{1} provides an overview of LR-FHSS and compares its goodput with the LoRa modulation. An analytical and simulation model was proposed to investigate the link budget and  Media Access Control (MAC) performance of LR-FHSS for DtS scenarios in~\cite{2}. The results obtained from these models confirm the feasibility of LR-FHSS DtS connectivity. The authors in~\cite{3} develop an outage probability analytical model for the DtS IoT network, considering factors such as channel fading, noise, and capture effect. The results reported in~\cite{1,2,3} reveal that LR-FHSS has a significantly higher capacity than the LoRa modulation. Furthermore, the LR-FHSS-enabled DtS network can offer connectivity services to numerous applications in remote areas, i.e., autonomous ships in the deep sea and smart farming in deserts~\cite{5}.

Despite the feasibility of DtS IoT, the limited energy resources at the terrestrial and space segment are one of the key challenges to satellite IoT~\cite{4}. The user devices envisioned for DtS IoT will be compact, low-cost, and powered by a battery that intends to last for years~\cite{5,9}. 
Therefore, it is important to understand the energy consumption characteristic of \textcolor{black}{terrestrial IoT devices}. The energy consumption of satellite IoT technology has not been studied extensively. Only a few studies examine the energy efficiency of IoT devices. \textcolor{black}{In~\cite{6}, the authors model and evaluate the energy performance of Iridium satellite IoT devices. The current consumption is measured using a RockBLOCK Mk2 device with an Iridium 9602 module and supports satellite IoT services. According to their findings, an Iridium IoT device powered by a 2400~mAh battery can maintain operations for an estimated 43.8~days by transmitting a 12~bytes message every 10 minutes. The works in~\cite{7,8} measure and model LoRa energy consumption for terrestrial networks. Specifically, in~\cite{7}, experiments have been conducted to measure the timing and current consumption of the different states (e.g., radio preparation, transmission, radio off, post-processing) involved in a LoRa transmission. The results also report the impact of LoRaWAN Data Rates (DR) settings on battery lifetime.}


In this paper, we carry measurements using LR-FHSS-enabled devices to identify and define the operational states of the system, their corresponding timings, and current consumption. These measurements are the foundation for our work, culminating by developing analytical models that accurately estimate the current consumption and ToA of LR-FHSS transmissions. To the authors' knowledge, no previous study has specifically investigated the LR-FHSS current consumption characteristics. The existing LR-FHSS literature~\cite{1,2,3,5,9,11,12} focuses primarily on radio propagation and network scalability theoretical aspects. Therefore, this paper aims to bridge this gap by providing empirical insights into the current consumption and ToA of LR-FHSS, complementing the existing theoretical understanding of this technology. The obtained results and developed models enable a more accurate study and characterization of the LR-FHSS devices' performance using analytical methods and simulators. \textcolor{black}{Our work can allow us to assess the feasibility of using this technology for different practical use cases with heterogeneous
connectivity needs, especially the satellite IoT.}

\vspace{-10pt}
\subsection{Our contributions}

The major contributions of this paper are as follows:

\begin{itemize}
    \item \textcolor{black}{we conduct an extensive state-of-the-art review and present the up-to-the-date background and technical details of LR-FHSS operations. Specifically, we accumulate the information from our experimental measurements, state-of-the-art publications~\cite{1,2,3,5,9,12}, LoRaWAN regional parameters document~\cite{RP}, and LR-FHSS device specifications~\cite{Datasheet1,USerManual} and the software~\cite{Github}.}
    
    \item we conduct measurements using the LR-FHSS-enabled LR1120 development kits~\cite{Datasheet1}. The measurement results from the DC Power Analyzer illustrate the key system modes the device goes through for transmitting an LR-FHSS packet. Furthermore, we examine how payload, transmit power, and LR-FHSS-based DR configurations influence these modes' timings and current consumption.
    
    \item we leverage measurement results to develop a more accurate analytical ToA model. Our analytical model closely aligns with the measurements, demonstrating a relative error lower than 0.3\%.
    
    \item we introduce an analytical model that estimates the current consumption of LR-FHSS. To the author's knowledge, our study is the first to present a current consumption model for LR-FHSS technology. We use the proposed models to theoretically estimate the battery lifetime of an LR-FHSS-enabled device.
    
    \item we make the measurement data and the script code used for data processing publicly available at GitHub [to be added later], allowing readers and researchers to access and utilize them for future studies and investigations.
\end{itemize}
\vspace{-10pt}
\subsection{Paper outline}
The rest of this paper is organized as follows. 
Section~\ref{sec:2} presents the background of LR-FHSS technology and the LR1120 development kit.
Section \ref{sec:3} discusses the experiment setup. 
Selected measurement results are presented in Section~\ref{sec:4}. 
The development of ToA and current consumption analytical models are detailed in Section~\ref{sec:5}. 
The key results derived from the analytical models and their comparison with our measurement results are discussed in Section~\ref{sec:6}.  
Finally, Section~\ref{sec:7} concludes this work with final remarks. 

\section{Technical background}
\label{sec:2}
\subsection{LR-FHSS}
\subsubsection{Physical layer}
The new LoRaWAN DRs (DRs 5-6 for the United States (US) and DRs 8-11 for the European Union (EU) region) feature LR-FHSS modulation~\cite{1,2}. LR-FHSS is based on Gaussian minimum-shift keying (GMSK) modulation, which uses bandwidth time (BT) products to define the pulse shapes. 
Currently, it features BT~=~1; however, the other values of BT are reserved for future use~\cite{USerManual}. BT~=~1 offers an instantaneous bit rate ${R_{b}}=$~488.28125~bit per second (bps)~\cite{Datasheet1}, including useful data as well as overhead, e.g., preamble, coding redundant, and cyclic redundancy check (CRC) bits.  Each coded bit (i.e., after forward error correction (FEC)) has a duration of approximately ${{T_{b}}~=~\frac{1}{R_{b}}}=$~2~ms, as explained in~\cite{3}.   To enhance network capacity and minimize collisions, LR-FHSS uses intra-packet frequency hopping. Besides the improved capacity, the high-frequency drift tolerance of 300~Hz/s (subject to 1.5~dB sensitivity degradation) makes LR-FHSS a prominent technology for DtS scenarios. A minimum signal-to-noise ratio (SNR) of 3.96 dB is required for successful LR-FHSS signal decoding~\cite{11}. With an occupied bandwidth (OBW) channel of 488 Hz, LR-FHSS offers a comparable link budget to LoRa DR0 operating at a 125 kHz bandwidth while delivering 200 times greater capacity.  \textcolor{black}{It features frequency hopping, low code rates, and header diversity, resulting in improved network scalability compared with conventional LoRa modulation.} Table~\ref{tab:tab1} compares the key characteristics of LR-FHSS and LoRa~\cite{LoRa}.

\begin{table}[t!]
\caption{Comparison of LoRa and LR-FHSS key features.} 
\centering
\begin{threeparttable}
\begin{tabular}{@{}cccc@{}}
\toprule
\textbf{}	& \textbf{LoRa} & \textbf{LR-FHSS}\\
\midrule
Modulation	& CSS	        & GMSK\\
Code rate & $\frac{4}{5}\dots\frac{4}{8}$	        &$\frac{1}{3},\frac{1}{2},\frac{2}{3},\frac{5}{6}$\\
Daily Maximum Capacity \tnote{1} & 1.2M uplinks/GW 	        &11M uplinks/GW\\
Frequency Drift Tolerance & 120 Hz/s 	        & 300 Hz/s\\
Uplink	& \checkmark 	        & \checkmark \\
Downlink	& \checkmark	        &$\Cross$\\
Spreading factors & \checkmark	        &$\Cross$\\
Header Diversity & $\Cross$        &\checkmark \\
Intra-packet hopping & $\Cross$	        &\checkmark\\
\bottomrule
\label{tab:tab1}
\end{tabular}
\begin{tablenotes}
   \item[1] Capacity in million (M) uplink packets per day per gateway for a 1.5~MHz channel bandwidth i.e., operating channel width (OCW)~\cite{Datasheet1}.  
  \end{tablenotes}

\end{threeparttable}
\end{table}

LR-FHSS uses convolution encoding to perform forward error detection and correction. Specifically, the header implies a fixed CR=$\frac{1}{2}$ while payload data fragments can use four code rate options CR=$[\frac{5}{6},\frac{2}{3},\frac{1}{2},\frac{1}{3}]$. However, current LoRaWAN specifications only support two code rates $\frac{1}{3}$ and $\frac{2}{3}$ reserving the remaining two for future use, as mentioned in the LR1120 user manual~\cite{USerManual}. Specifically, in ETSI region, DR8/DR10 and DR9/DR11 imply coding with the code rate equivalent to $\frac{1}{3}$ and $\frac{2}{3}$, respectively. Similarly, in FCC operational areas, DR5 and DR6 imply coding with the code rate equivalent to $\frac{1}{3}$ and $\frac{2}{3}$, respectively. Notably, the lower code rate improves the gateway’s ability to correctly demodulate the packet in the presence of noise and interference.

\subsubsection{Frame structure}
\begin{figure}[t!]
	\begin{center}
		\includegraphics*[width=0.5\textwidth]{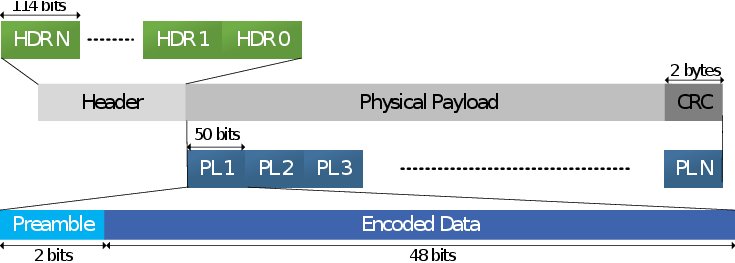}
	\end{center}
 \vspace{-10pt}
	\caption{The key components of the LR-FHSS packet structure.} 
	\label{fig1}
 
 \vspace{-15pt}
\end{figure}

Fig. \ref{fig1} illustrates the LR-FHSS frame structure. LR-FHSS divides the payload into small fragments and transmits over multiple OBW channels following the frequency hopping sequence picked by a pseudo-random number generator. The packet header contains essential information to notify the gateway about the frequencies and hopping sequence of the payload fragments~\cite{12}. Unlike LoRa, the LR-FHSS device transmits several replicas of headers (${N_{H}}=$~1 \dots 4), where the DR setting defines the number of replicas. A gateway should receive at least one of the ${N_{H}}$ transmitted headers to decode a packet successfully. However, header diversity improves immunity against co-channel interference, which costs longer ToA. In LoRaWAN protocol, DR5 (in FCC region) and DR8/DR10 (in ETSI region) imply ${N_{H}}=$~3, while DR6 (FCC) and DR9/DR11 (ETSI) feature ${N_{H}}=$~2~\cite{1,2}.

Each encoded header comprises ${H_{b}}=$~114~bits resulting in header duration as ${T_{H}}=\frac{{H_{b}}}{{R_{b}}}=$~233.472~ms. Contrary, each encoded payload (payload plus CRC) fragment has ${F_{b}}=$~50~bits containing 48~payload~bits and 2~bits preamble in each hop~\cite{USerManual,3,11}. Therefore, each payload fragment is ${T_{P}}=\frac{F_{b}}{R_{b}}=$~102.4~ms long. However, the last payload fragment can be shorter.
\if{0}
According to~\cite{USerManual}, LR-FHSS can support a payload of up to 255 coded bytes. However, strict limitations apply, and the user payload length varies with header replicas and the code rates as shown in Table~\ref{tab:tab2}. 
\begin{table}[t!]
\caption{Maximum allowed user payload (in bytes) of LR-FHSS~\cite{USerManual}.} 
\centering
\begin{threeparttable}
\begin{tabular}{@{}cccccc@{}}
\toprule
\textbf{Code Rate}	& & \textbf{   
 Header replicas}  & & \\
\textbf{}	& \textbf{1} & \textbf{2} & \textbf{3} & \textbf{4}\\
\midrule
$\frac{5}{6}$	& 189	        & 178 & 167	        & 155\\
$\frac{2}{3}$	& 151	        & 142 & 133	        & 123\\
$\frac{1}{2}$	& 112	        & 105 & 99	        & 92\\
$\frac{1}{3}$	& 74	        & 69 & 65	        & 60\\
\bottomrule
\label{tab:tab2}
\vspace{-15pt}
\end{tabular}
\end{threeparttable}
\end{table}
\fi
\subsubsection{Channels and Frequency hopping}
 
\begin{figure}[t!]
	\begin{center}
		\includegraphics*[width=0.5\textwidth]{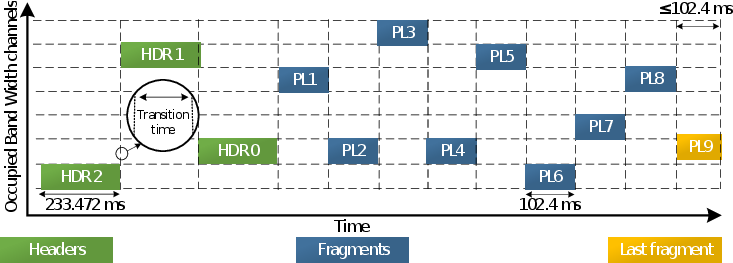}
	\end{center}
 \vspace{-10pt}
	\caption{Illustrative frequency hopping profile of a single LR-FHSS packet of payload $L = 15$ bytes and DR8 featuring code rate =$\frac{1}{3}$ and $N_{H} = 3$} 
	\label{fig2}

 \vspace{-15pt}
\end{figure}

LR-FHSS provides a range of configuration options for channel bandwidths, hopping grid size, and minimum frequency separation. The whole frequency band is divided into multiple operating channel width (OCW) channels, allowing the cumulative bandwidth to span from 39.06~kHz to 1.5742~MHz, depending on the local frequency regulations~\cite{Datasheet1}. For frequency hopping, a single LR-FHSS OCW channel is further divided into multiple subchannels with a bandwidth of 488~Hz, named OBW channels. At any moment, a single device can transmit its data (header or fragment) only in one OBW channel, also known as a hop. Note that multiple OBW channels are required to transmit the full LR-FHSS packet. Fig.\ref{fig2} depicts the transmission of 15~bytes payload implying $\text{CR}=\frac{1}{3}$, header replica ${N_{H}}=$~3 resulting in number of payload fragments ${N_{F}}=$~9.

\subsubsection{Existing Time-on-Air models}
The existing LR-FHSS ToA models are based primarily on LoRaWAN regional specification document~\cite{RP}.
\subsubsection*{Model I} Departing from the information in~\cite{2}, the ToA of an LR-FHSS packet for payload $L$ bytes is expressed as

\begin{align}
T_{model\_I}\!=\!  N_{H}T_{H} + T_{P}\left\lceil \frac{L+3}{M} \right\rceil,
\label{m1}
\end{align}
where $M$ equals 2 and 4 for DR8/DR10 and
DR9/DR11, respectively. This model implies all the payload fragments have the same duration.
\subsubsection*{Model II}
Unlike the previous model, the work in~\cite{3,11} accounts for the fact that the last fragment could be shorter than standard ${T_{H}}=$ 102.4~ms.  Following this model, the ToA can be calculated as
\begin{align}
T_{model\_II} &= N_{H}T_{H} + N_{PL}T_{P},
\label{m2}
\end{align}
where $N_{PL}$ is the number of payload (data payload plus CRC) fragments after FEC coding. Since we know each hop can accommodate 48~bits (6~bytes), thus, $N_{PL}$ becomes
\begin{align}
N_{PL} &= \frac{(L + P_{\text{CRC}})}{6\text{CR}},
\end{align}
where $P_{CRC}=$2~bytes accounts for CRC.

\subsection{LR1120 development kit}
\label{LR1120}
The LR1120 is the first commercially available development kit featuring an LR-FHSS modulator. This real transceiver allows us to get insight into the LR-FHSS operation, system modes, their duration, and associated current consumption.  Therefore, in what follows, we briefly introduce the LR1120 modem and respective development kit, which we will further use in our experimental campaign.
\if{0}
\subsubsection{Radio configuration}
Upon powering on, the LR1120 device initiates an automatic calibration procedure and enters \textit{standby mode}, the default operational mode. To transmit a packet, the LR-FHSS-enabled LR1120 device undergoes a series of operations in the following sequential order \cite{USerManual}: 
\begin{enumerate}
    \item the device configures the carrier frequency, transmit power, radio frequency (RF) switch, power amplifiers (PA), and over-current protection mechanism; 
    \item it sets modulation to LR-FHSS;
    \item it builds the LR-FHSS frame following the assigned DR (header replica and CR) and bandwidth; this stage also includes Viterbi encoding;
    \item LR1120 sets the circuits to transmitter mode and starts the LR-FHSS transmissions; and finally,
    \item it waits for an interrupt that indicates the completion of the transmission. Upon receiving the interruption, it turns off the transmitter. 
\end{enumerate}
\fi
\subsubsection{Power amplifiers}
The LR1120 device includes a low-power amplifier (LPA) and high-power amplifier (HPA) to facilitate its sub-GHz operations. The selection between these amplifiers depends on the configured transmit power ($P_{tx}$). The LPA is primarily optimized for $P_{tx}$ below +14~dBm, while it can also support +15~dBm. In contrast, the HPA is specifically designed for generating up to $P_{tx}=$+22~dBm~\cite{USerManual}.

\subsubsection{System modes}

\begin{figure*}[t!]
	\begin{center}
		\includegraphics[height=1.8in]{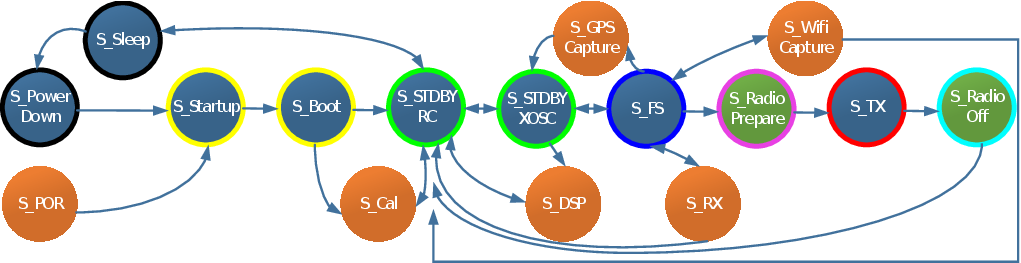}
	\end{center}
 \vspace{-10pt}
	\caption{LR1120 operational modes, where the states marked by filled green circles are not reported in the user manual \cite{USerManual}. We identify these green states from experimental measurements and analyse the code of the transceiver's firmware, while the orange ones are irrelevant in this work. Notably, colour borders distinguish between the different states and build a relation with Fig.~\ref{fig7}}. 
	\label{fig3}
 \vspace{-15pt}
\end{figure*}

Fig.~\ref{fig3} illustrates the LR1120 operational modes. The \textit{power down} mode has the lowest power consumption since it stops all clocks and does not retain any data. Similarly, the \textit{sleep} mode offers a low power consumption option. However, it retains the configuration register values and stores the firmware data in random-access memory (RAM).

Upon \textit{startup} (S\_Startup), the \textit{bootloader} performs firmware validation before transitioning to the \textit{standby} state (S\_STBY\_RC), the default mode of LR1120. The \textit{standby} modes, S\_STBY\_RC and S\_STBY\_XOSC,  differ in clock oscillator selection, with the former utilizing the internal Resistance-Capacitance (RC) oscillator at 32 MHz. At the same time, the latter employs an external crystal clock, allowing faster transitions to other modes. To transmit a packet, the device leaves the \textit{standby} mode and enters \textit{frequency synthesis} (S\_FS) mode to activate the Phase-Locked Loop and associated regulators.

The LR1120 user manual (see Figure 2-1 in~\cite{USerManual}) does not present a \textit{radio preparation} mode. However, our measurements indicate that the device enters a \textit{radio preparation} (S\_Radio\_Prepare) state for around 100 milliseconds before \textit{transmission} (S\_TX) mode. Next, in S\_TX mode, the LR1120 transmits the LR-FHSS packet at the configured carrier frequency and DR using the relevant power amplifier to achieve the desired output power. After completing the transmission, the device enters a \textit{radio off} (S\_Radio\_Off) mode, deactivating the power amplifier and regulators. It's worth noting that this mode is not documented in the LR1120 state diagram (see Figure 2-1 in~\cite{USerManual}). Finally, the device reverts to the default \textit{standby} mode (S\_STBY\_RC). However, the code we use for testing specifically configures the device to enter \textit{sleep} (S\_Sleep) mode and remain there until the next transmission.

\section{Experimental Setup}
\label{sec:3}
\begin{figure*}[t!]
	\begin{center}
		\includegraphics*[width=1\textwidth]{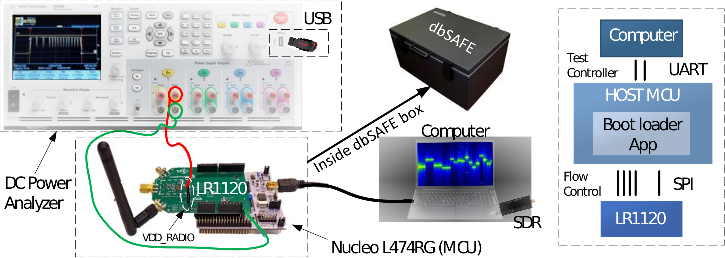}
	\end{center}
 \vspace{-10pt}
	\caption{The structural diagram of the experimental setup for current consumption measurements of LR-FHSS transmissions.} 
	\label{fig4}
\end{figure*}
\subsection{Hardware}
This work uses LR1120 development kits as a reference LR-FHSS transmitter for our measurement~\cite{USerManual}. \textcolor{black}{Notably, our current consumption and ToA model will remain valid for other LR-FHSS devices. However, different power amplifier configurations could impact the current consumption in \textit{transmission} (S\_TX) mode. Any such differences can be incorporated into our model through minor adjustments, which can be done later depending on the availability of LR-FHSS new device models.}

Fig.~\ref{fig4} shows the structural diagram of our experimental setup. One can see that LR-FHSS-enabled LR1120 is further connected to the ST Microelectronic NUCLEO-L476RG, which serves as a host Microcontroller Unit (MCU). We installed software on the host MCU responsible for programming the LR1120. MCU and LR1120 radio communication based on Serial Peripheral Interface (SPI)~\cite{SPI}.  
For measuring the current consumption, duration of each operation state, and ToA, we utilized the Agilent N6705B DC Power Analyzer. Specifically, we performed the following steps:
\begin{itemize}
  \item To ensure accurate measurements and minimize energy consumption and fluctuations, we removed the light-emitting diode (LED) (LD1) from the MCU. Subsequently, we mounted the LR1120 onto the MCU. We established a connection between the MCU and a computer to facilitate the installation of the LR-FHSS software and overall control over the experiment.
   \item Next, to monitor the current consumption of the LR-FHSS-enabled LR1120, we removed the VDD\_RADIO jumper and directly connected VDD\_RADIO to the positive output (red wire in Fig.~\ref{fig4}) of a DC Power Analyzer. Simultaneously, we used the MCU and LR1120 radio ground (GND) as a common reference and connected it to the negative output (green wire in Fig.~\ref{fig4}) of the DC Power Analyzer. We utilized a 3.3~V DC supply for the measurements, with a maximum current limit of 180~mA current and a sampling period of 20.48~microseconds ($\mu$s). The MCU was directly powered from a computer using a mini USB connector cable. This setup exclusively enables the current consumption and timing measurement of different LR-FHSS modes without considering the current consumption of the host MCU.
  \item In the third step, we placed the testbed inside a dbSAFE box to ensure RF isolation.
    \item Finally, we connected and configured the Nooelec RTL-Software Defined Radio (SDR) to a computer, enabling us to observe and visualize the frequency spectrum of the LR-FHSS transmissions.
\end{itemize}
\if{0}
\begin{table}[t!]
\caption{The list of experimental testbed hardware and purpose.} 
\centering
\begin{threeparttable}
\begin{tabular}{@{}lll@{}}
\toprule
\textbf{Devices name}	& \textbf{Model} & \textbf{Purpose}\\
\midrule
Development Kit	& LR1120	        & LR-FHSS transmissions\\

DC Power Analyzer & Agilent N6705B & Current measurements\\
Microcontroller & Nucleo-L476RG & Host controller\\
Shielded box & dbSAFE & RF isolation\\
Software Defined Radio & Nooelec RTL-SDR & Spectrum analyzer\\
\bottomrule
\label{tab:tab3}

\vspace{-15pt}
\end{tabular}
\end{threeparttable}
\end{table}
\fi
\vspace{-10pt}
\subsection{Software}
We program the MCU using the LR11xx radio drivers published by Semtech on GitHub~\cite{Github}. It's important to note that this driver does not implement LR-FHSS reception. We use the Keil IDE to introduce changes and build the software, specifically implementing three modifications in the code:
\begin{itemize}
  \item Firstly, in \verb|LR1110_lr_fhss_ping.c| file, we disabled the TX and RX LEDs of LR1120 to prevent measurement fluctuations and additional current consumption.
  \item Secondly, by introducing changes to the code, we configure the radio to transit into \textit{sleep} mode after transmitting a packet, even though the default mode was standby mode. Specifically, we modify the \verb|enter_standby_then_sleep_mode()| function in the \verb|LR1110_lr_fhss_ping.c| file.
  \item Third, we update the \verb|LR1110_lr_fhss_ping.c| file to enable the following three measurement scenarios
  \begin{enumerate}[label=\roman*)]
      \item 10~bytes constant payload size and variable $P_{tx}$ ranging from 0 to 22~dBm; 
      \item constant $P_{tx}$ 14~dBm and variable payload 10 to 65~bytes; 
      \item repeating the measurements above for DR8 and DR9.
  \end{enumerate}
\end{itemize}
Lastly, we developed a MATLAB script for processing the collected data with the following functionalities: 
\begin{enumerate}[label=\roman*)]
\item allows identification of different transition states involved in LR-FHSS transmission, 
\item  calculates duration and current consumption for each state, and 
\item  total ToA calculation of \textit{transmission} (S\_TX) mode.
\end{enumerate}
\section{Measurement results}
\label{sec:4}
\begin{figure}[t!]
	\begin{center}
		\includegraphics*[width=0.5\textwidth]{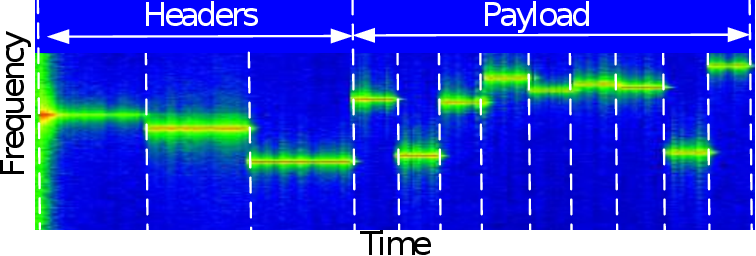}
	\end{center}
\vspace{-10pt}	\caption{Frequency spectrum of LR-FHSS illustrating intra-packet frequency hopping profile of a single LR-FHSS packet of $L =$ 15-byte payload and DR8 featuring code rate =$\frac{1}{3}$ and $N_{H} = 3$.} 
	\label{fig5}

 \vspace{-15pt}
\end{figure}

A series of test measurements were initially conducted to ensure the accuracy and reliability of the hardware and software configuration. Fig.~\ref{fig5} depicts the LR-FHSS packet spectrum captured by CubicSDR software, visually representing the intra-packet frequency hopping. This validation step confirms the proper functioning and appropriateness of the implemented hardware and software setup.

We perform several measurements to characterize LR-FHSS average current consumption and ToA.  To accomplish the former, we conduct twenty measurements under varying $P_{tx}$ levels, ranging from 0~dBm to 22~dBm, specifically for DR8 and DR9. To develop a realistic ToA model, we carry out measurements across a range of variable payload sizes, spanning from 10 to 65 bytes, for both LR-FHSS DRs, i.e., DR8 and DR9. This accounts for different code rates and header diversity configurations of the DRs.
\begin{figure}[t!]
	\begin{center}
		\includegraphics*[width=0.5\textwidth]{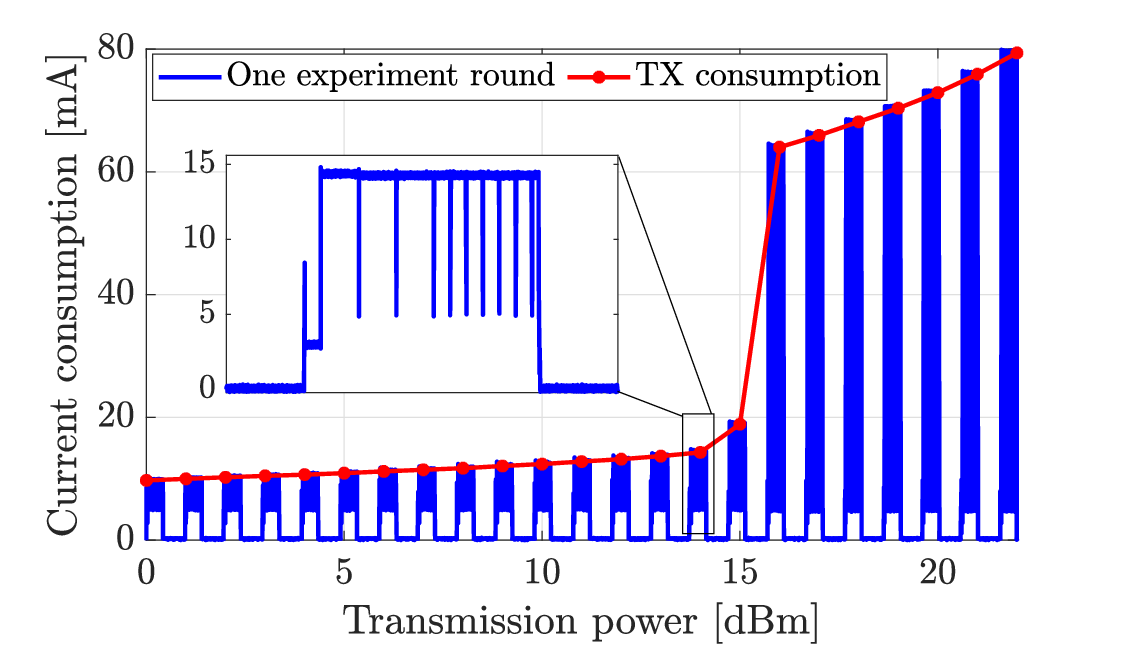}
	\end{center}
 \vspace{-10pt}
	\caption{The impact of variable transmit power on LR-FHSS radio current consumption for one experimental round. We have conducted twenty rounds of measurements, resulting in over 460 packet transmissions for each DR.}
	\label{fig6}

 \vspace{-15pt}
\end{figure}
Fig.~\ref{fig6} shows the current consumption for just one out of twenty experimental rounds. One can see that the current consumption increases significantly after $P_{tx}=$ 14~dBm when the power amplifier switches from LPA to HPA. Fig.~\ref{fig7} illustrates the current consumption profile of a single LR-FHSS transmission. The transmission undergoes different states, as mentioned in Section~\ref{LR1120}. Each state within the transmission process reveals distinct characteristics regarding duration and associated current consumption. Table~\ref{tab:tab4} summarizes these states, their duration, and the average current consumption computed over all 20 experiment runs.
\begin{figure*}[t!]
	\begin{center}
		\includegraphics*[width=0.95\textwidth]{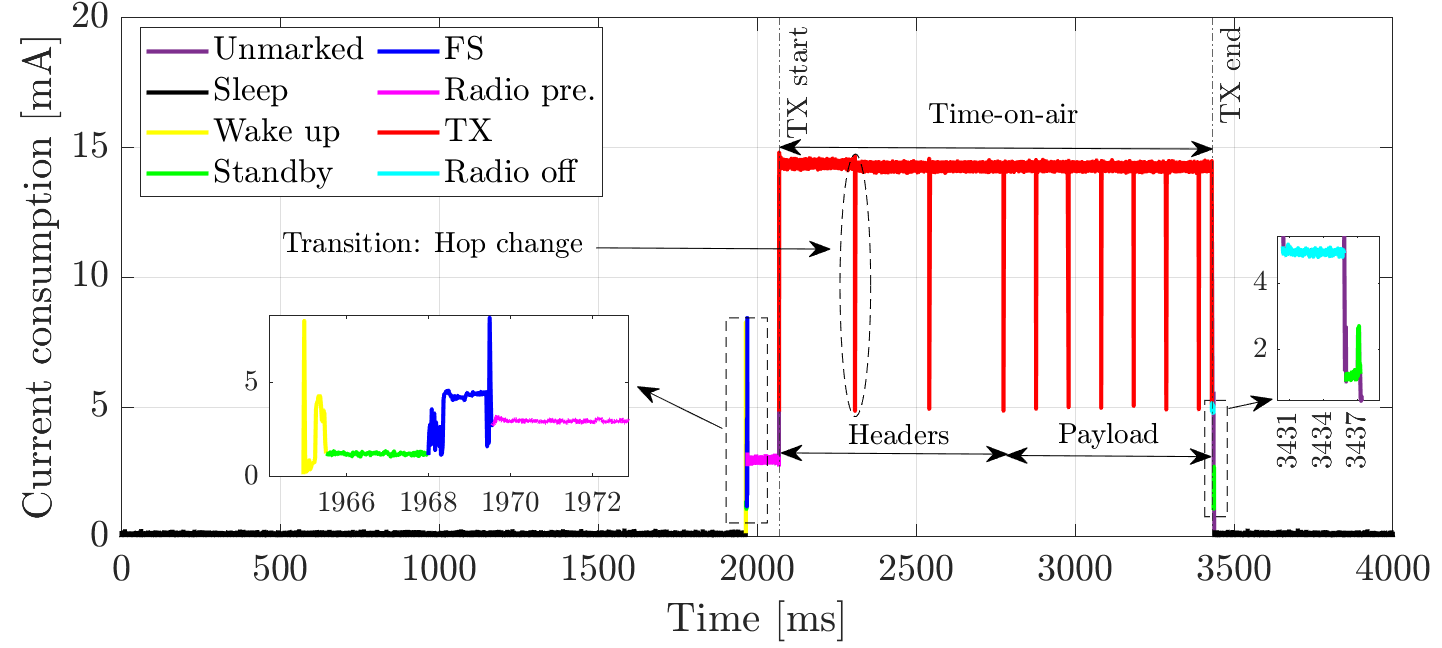}
	\end{center}
 \vspace{-10pt}
	\caption{LR-FHSS current consumption for an illustrative packet transmission with different states of transmissions color-marked according to the operational modes shown in Fig.~\ref{fig3}.} 
	\label{fig7}
  \vspace{-10pt}
\end{figure*}

\begin{figure}[t!]
	\begin{center}
		\includegraphics*[width=0.5\textwidth]{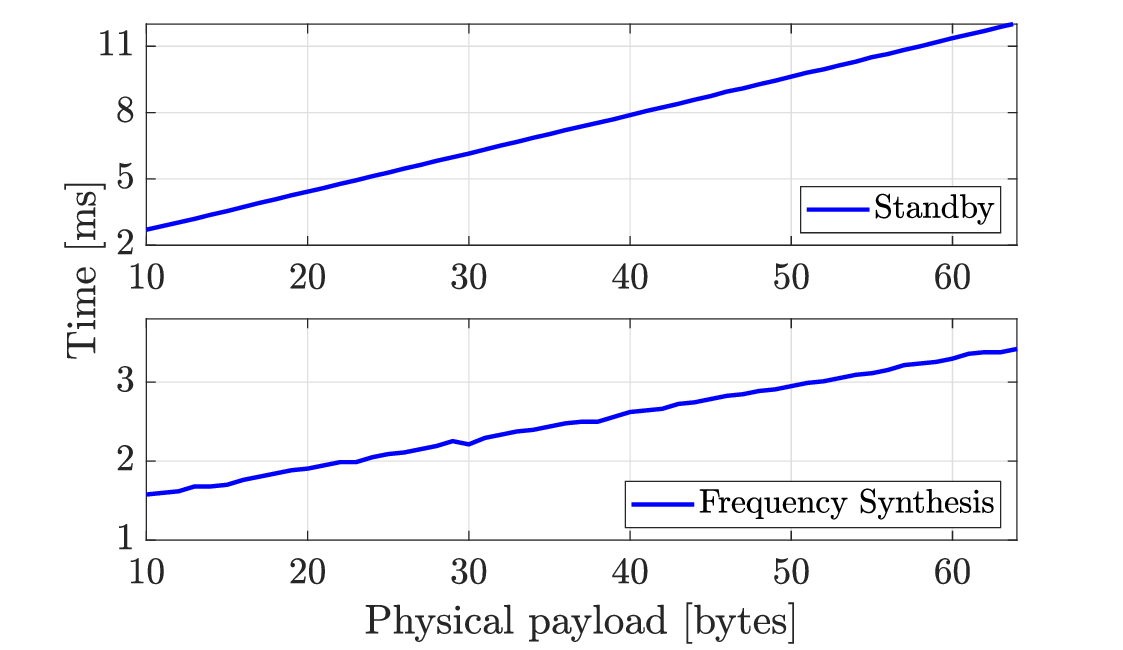}
	\end{center}
  \vspace{-10pt}
	\caption{The impact of payload size on the timings of \textit{standby} and \textit{frequency synthesis} mode.} 
	\label{fig8}
  \vspace{-10pt}
\end{figure}

Specifically, in Fig.~\ref{fig7}, one can see that initially, the device is in \textit{sleep} mode, which has the lowest current consumption of $I_{sleep}=$~0.053~mA. To transmit a packet, the device \textit{wakes up} (State 1) and enters the default \textit{standby} mode (State 2). Notably, the duration of this state depends on the payload, as reported in Fig.~\ref{fig8}. However, it is worth mentioning that the average current consumption in the \textit{standby} state remains constant $I_{std}=$ 1.229~mA, irrespective of the payload size, DRs, and $P_{tx}$ levels. Subsequently, the device engages in \textit{frequency synthesis} named State 3. Similar to the \textit{standby} mode (State 2), the duration of \textit{frequency synthesis} mode depends on the payload size as shown in Fig.~\ref{fig8}. 

\begin{table}[t!]
\caption{LR-FHSS states during packet transmission, their duration, and average current consumption} 
\centering
\begin{tabular}{cllclc}
\hline
State & State & \thead{Duration $T_{i}$} & &\thead{Current $I_{i}$} &\\
\cline{3-6}
 number $i$ & name &  sym & ms & sym & mA \\
\hline
1 & Wake up & $T_{wu}$ &0.4301&$I_{wu}$&1.9\\
2 & Standby & $T_{std}$ &Fig.~\ref{fig8}&$I_{std}$ & 1.229\\
3 & FS & $T_{fs}$ &Fig.~\ref{fig8}&$I_{fs}$&3.7392\\
4 & Radio Pre. & $T_{pre}$ &99.67&$I_{pre}$ & 2.968 \\
5 & Transmission & $T_{tx}$ & (\ref{Eq:ToA}) &$I_{tx}$ &Fig.~\ref{fig6}\\
6 & Radio off & $T_{off}$ &9.45&$I_{off}$ & 4.94\\
7 & Standby & $T_{std}^{\prime}$ &1.044&$I_{std}^{\prime}$ & 1.229\\
8 & Sleep & $T_{sleep}$ &   (\ref{Eq:sleep})&$I_{sleep}$ & 0.053\\
\hline
\end{tabular}
\label{tab:tab4}

\vspace{-15pt}
\end{table}

In the next phase, the LR1120 device prepares the radio packet for transmission. Our measurement results reveal that the duration and current consumption of the \textit{radio preparation} mode (State 4) remains consistent around $T_{pre}=$~99.67~ms $I_{pre}=$~2.968~mA across all configurations.

\begin{figure}[t!]
	\begin{center}
		\includegraphics*[width=0.5\textwidth]{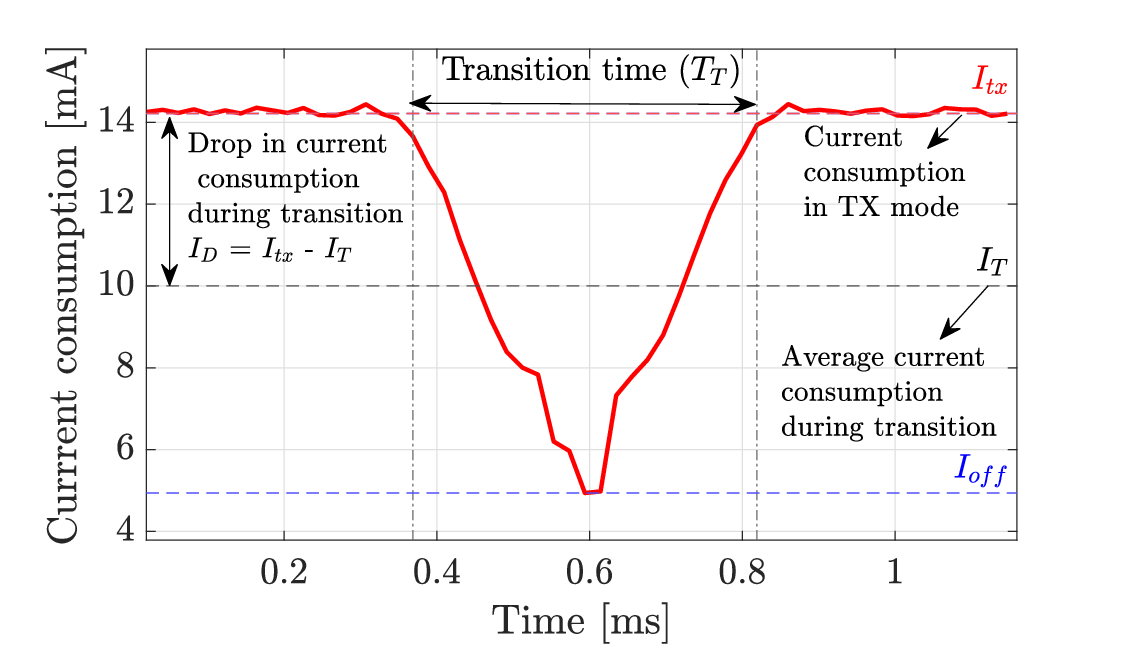}
	\end{center}
  \vspace{-10pt}
	\caption{The transition time between the hop change during the intra-packet frequency hopping.} 
	\label{fig9}
  \vspace{-15pt}
\end{figure}

Following it, the device initiates the packet \textit{transmission} (State 5), which has the highest current consumption compared to other states. The current consumption is directly influenced by the selected transmit power level. 
The duration of the \textit{transmission} state corresponds to the packet ToA, which is influenced by the payload size and the chosen DRs. From the measurements, we identify LR-FHSS exhibits an average transition time $T_{T}=$~0.61~ms to switch the OBW channel for the next hop, as illustrated in Fig.~\ref{fig9}. However, it is important to note that the total transition time experienced during packet transmission depends on the number of header replicas and payload fragments. To accurately model the LR-FHSS ToA, it is crucial to consider the total transition time.

Upon completing the transmission, the device transitions to the \textit{radio off} mode (State 6) and returns to the default \textit{standby} mode (State 7). Unlike the initial \textit{standby} mode (State 2), the duration of the final \textit{standby} state remains consistent across different configurations and payload sizes. Finally, the device returns to the \textit{sleep} (State 8). Unlike LoRa, LR-FHSS is designed to support uplink transmissions only. Therefore, in our experiments, the receiving windows were disabled, and our measurement does not account for the current consumption in the receive mode. We focus on analyzing and modeling the current consumption during the transmit process of LR-FHSS modulation. Notably, the LR-FHSS-enabled devices can use LoRa modulation for downlink communication. Therefore, existing LoRa ToA and energy consumption models~\cite{7,8} can be used for the downlink analysis.

\vspace{-5pt}
\section{Proposed analytical Models}
\label{sec:5}
\subsection{LR-FHSS Time-on-Air}
Our research is grounded in real LR-FHSS measurement results presented in Section~\ref{sec:4} and references to Semtech documents~\cite{Datasheet1} and the LR1120 software~\cite{USerManual}. We leverage this information to develop an accurate analytical model for LR-FHSS ToA.  We consider ToA as the total time required to transmit all the encoded bits, including header, payload, preambles, and CRC at a given bit rate ${R_{b}}$. Additionally, we account for the transition time ${T_{T}}$ required to switch the OBW channels during frequency hopping as illustrated in Fig.~\ref{fig9}. The proposed model takes into account these facts to calculate the total ToA as
\begin{align}
T_{tx} &= \frac{P_{B}}{R_{b}} + T_{T}N_{T},
\label{Eq:ToA}
\end{align}
where $N_{T} = N_{H} + N_{F} -1$ is the number of transitions to change OBW channels for intra-packet frequency hopping, and ${P_{B}}$ is the total number of encoded bits in an LR-FHSS packet calculated as
\begin{align}
P_{B} &= H_{b} N_{H} + P_{L},
\end{align}
where ${P_{L}}$ represents the total number of encoded payload bits in the LR-FHSS packet, including the preamble ($P_{b}=$ 2 bits) for each payload fragment as
\begin{align}
P_{L} = P_{L}^{\prime} + P_{b}N_{F}
\end{align}
where $P_{L}^{\prime}$ denotes the number of payloads plus CRC bits after the FEC coding, i.e., 
\begin{align}
P_{L}^{\prime} &=  \frac{8(L + P_{CRC})}{\text{CR}} + O_{B}
\end{align}
where $L$ is the number of uncoded physical layer payload in bytes, and ${O_{B}}=$ 6 denotes the number of overhead bits defined in LR-FHSS software~\cite{Github}. Finally, the number of payload fragments $N_{F}$ in an LR-FHSS packet is given by
\begin{align}
N_{F} &=  \biggl \lceil \frac{P_{L}^{\prime}}{48}  \biggr \rceil 
\end{align}
\vspace{-15pt}
\subsection{LR-FHSS Current Consumption}
\textcolor{black}{LR-FHSS uses multiple OBW channels for intra-packet frequency hopping during a single transmission. Each transition between OBW channels, known as a hopping shift, requires a transition time ($T_{T}$).} It is important to note that the current consumption undergoes a substantial reduction during these transitions, reaching as low as $I_{off}=$ 4.94 mA, as demonstrated in Fig. \ref{fig9}. We account for this aspect in our work to accurately model the current consumption behavior of LR-FHSS transmissions.

Let $N_{\text{states}}=$ 8 be the number of states, $T_{i}$ and $I_{i}$ represent the duration and current consumption of these states as given in Table~\ref{tab:tab4}. Thus, the total average current consumption becomes
\begin{align}
\overline{I_{\text{avg}}} &= \frac{T_{\text{active}}.I_{\text{active}} + T_\text{{sleep}}.I_{\text{sleep}}}{T_{\text{notification}}},
 \label{eq:9}
\end{align}
\begin{align}
\overline{I_{\text{avg}}} &= \frac{1}{T_{\text{notification}}} \Biggl(\sum_{i=1}^{N_{\text{states}}} T_{i}.I_{i} - T_{T}.I_{D}.N_{T}\Bigg), 
 \label{eq:10}
\end{align}
\textcolor{black}{where $I_{D}$ account for the total drop in current consumption during the OBW channel changes for intra-packet frequency hopping as}
\begin{align}
I_{D} =  I_{tx}- \overline{I_{T}} ,
 \label{eq:11}
 \end{align}
where $\overline{I_{T}}$ is the average current consumption during the transition time $T_{T}$. In Fig.~\ref{fig9}, the drop in current from $I_{tx}$ to $I_{off}$ and then raise in current from $I_{off}$ to $I_{tx}$ exhibit a triangle shape. Using the triangle centroid formula, the equation for the average current consumption $\overline{I_{T}}$ can be expressed as 
\begin{align}
 \overline{I_{T}} =  \frac{I_{tx} + I_{off} + I_{tx}}{3}
 \label{eq:12}
 \end{align}

In (\ref{eq:9}) and (\ref{eq:10}), $T_{\text{notification}}$ is the reporting period, the time between two consecutive periodic transmissions calculated as
\begin{align}
T_{\text{notification}} = T_{\text{sleep}} +  T_{\text{active}},  
\label{Eq:sleep}
\end{align}
where $T_{\text{active}}$ represents the total duration of all active states combined.
\begin{align}
T_{\text{active}}\!=\!T_{wu} \!+\! T_{std}\!+\! T_{fs}\!+ \!T_{pre}\!+\!T_{tx}\!+\!T_{off}\!+\!T_{std}^{\prime}.  
 \label{eq:14a}
\end{align}
Next, these results can be used to estimate, e.g., a theoretical lifetime of operation from the battery. For example, if to imply a linear battery model, the lifetime is given by:
\begin{align}
T_{\text{lifetime}} &= \frac{C_{\text{battery}}}{\overline{I_{\text{avg}}}}, 
 \label{eq:14}
\end{align}
where $C_{\text{battery}}$ is battery capability expressed in mAh. \textcolor{black}{ Note that \eqref{eq:14} estimates the maximum lifespan of an ideal battery; however, real-world batteries degrade and experience a decline in their rated capacity over time.}
\vspace{-10pt}
\section{Discussion}
\label{sec:6}

\begin{figure}[t!]
	\begin{center}
		\includegraphics*[width=0.5\textwidth]{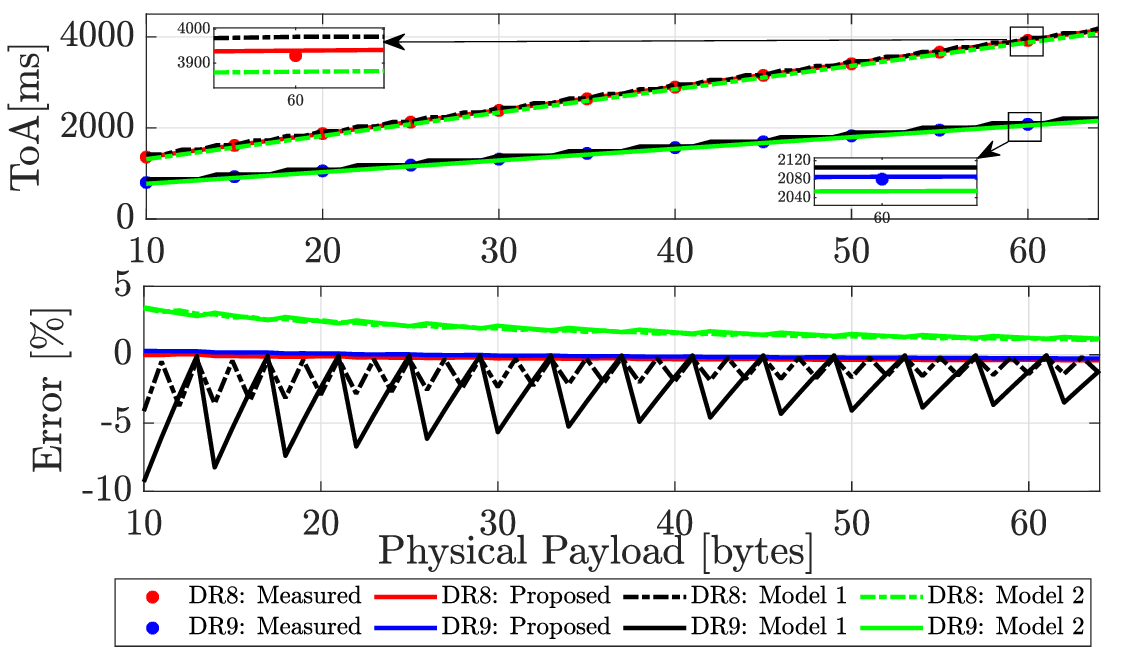}
	\end{center}
  \vspace{-10pt}
	\caption{Top section of the figure compares the ToA results obtained from measurements, the proposed model, and the state-of-the-art models~\cite{RP,3,11} as a function of variable payload. The bottom part of the figure reveals the error percentage when compared to the measurement ToA.}
	\label{fig10}
  \vspace{-10pt}
\end{figure}

Fig~\ref{fig10} shows the LR-FHSS ToA as a function of variable payload. The ToA values obtained from our proposed analytical model closely align with the experimental measurements, confirming the validity of our model. However, the existing state-of-the-art \textit{Model I}~\cite{RP} reports up to 55~ms higher ToA than the results obtained from our analytical model and measured data. Conversely, \textit{Model II}~\cite{3,11} gives a ToA approximately 47~ms lower.

The lower section of Fig.~\ref{fig10} reveals the relative error of ToA from different models relative to the measurement results. One can see that the proposed analytical ToA model features the lowest relative error, which is only 0.3\%. For 10 bytes payload, \textit{Model I and Model II} demonstrate the maximum error of -9.2\%  and 3.4\%, respectively. \textcolor{black}{These differences and fluctuations are mainly due to the mathematical formulation of \eqref{m1} and \eqref{m2}, specifically, the \textit{Model I} depicts sawtooth behavior because \eqref{m1} does not account for the fact that the last payload fragment could be shorter than 102.4 ms. Thus, this model introduces significant inaccuracy.}  Notably, the relative maximum error gradually decreases with the increase of the payload size for all the models. Our proposed approach reveals the best performance among these models, demonstrating the smallest relative error as we account for transition time $T_{T}$.

\begin{figure}[t!]
	\begin{center}
		\includegraphics*[width=0.5\textwidth]{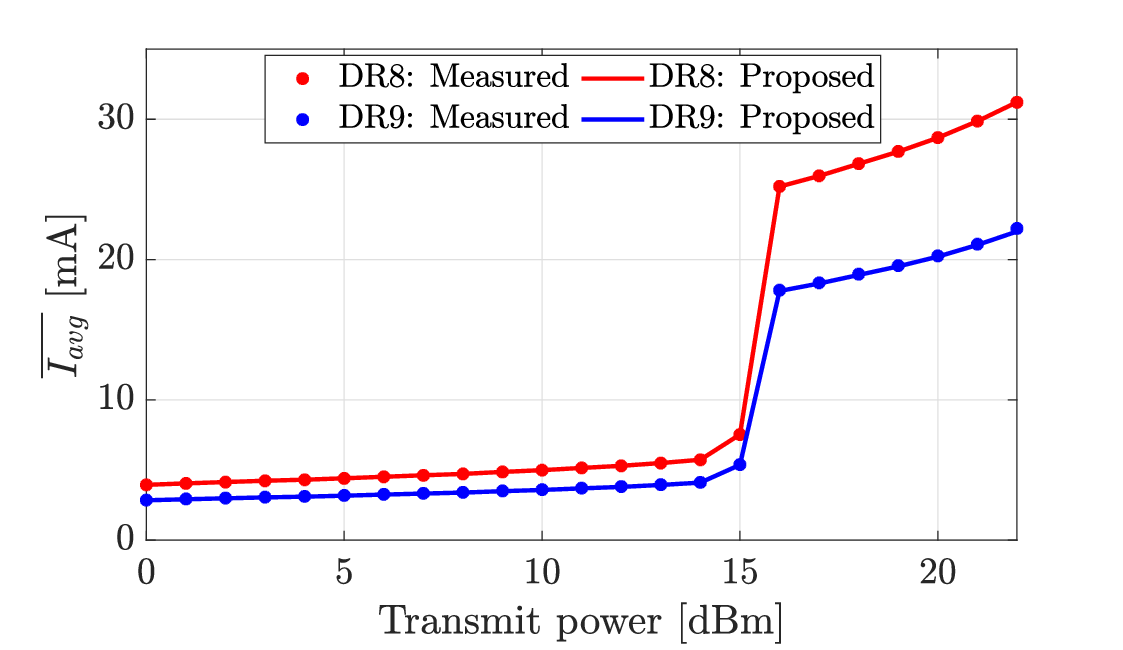}
	\end{center}
	 \vspace{-10pt}\caption{Comparison of measured and analytical average current consumption ($\overline{I_{\text{avg}}}$) for single packet transmission as a function of transmit power ($P_{tx}$).} 
	\label{fig11}
  \vspace{-10pt}
\end{figure}

In Fig.~\ref{fig11}, we compare the average current consumption $\overline{I_{\text{avg}}}$ during one packet transmission obtained from the measurement and our analytical model as a function of variable transmit power. The results reveal a strong correlation, thereby confirming the accuracy of our LR-FHSS current consumption model. To our knowledge, no available research or literature presents a current consumption model for LR-FHSS. Therefore, we can not compare our results to any other model. 

\begin{figure}[t!]
	\begin{center}
		\includegraphics*[width=0.5\textwidth]{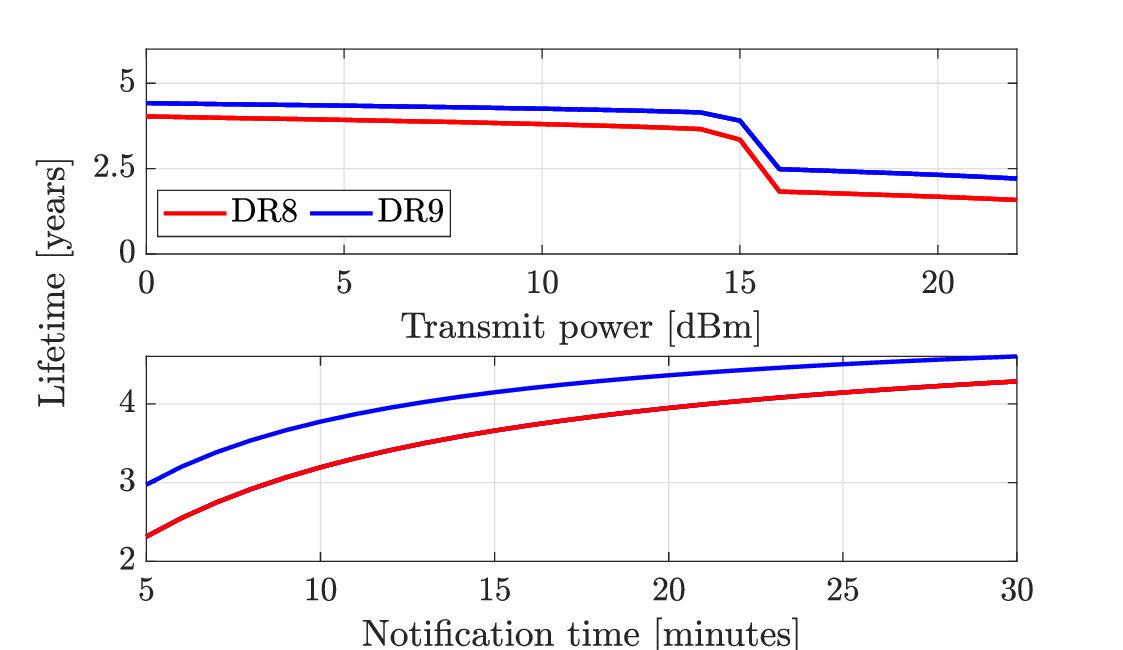}
	\end{center}
  \vspace{-10pt}
	\caption{The subplot at the top presents the battery theoretical lifetime for 10 bytes LR-FHSS transmission as a function of transmit power and notification time. The lower subplot implies a constant 14 dBm transmit power.} 
	\label{fig12}
  \vspace{-10pt}
\end{figure}

Finally, we theoretically evaluate the battery life of an LR-FHSS-enabled device. In the upper chart of Fig.~\ref{fig12}, we present the theoretical battery life as a function of variable transmission power. We imply the device transmits 10-byte packets every 15 minutes.  DR9 demonstrates a better lifetime owing to lower ToA. For both DRs, the battery lifetimes remain above 3.6~years up to a transmit power of 14~dBm. However, when the transmit power exceeds 14~dBm, and the device begins operating on HPA, the battery life gradually decreases to almost half - nearly 2.2 and 1.58~years for DR8 and DR9, respectively, at the maximum transmit power.

More than this, the lower chart of Fig.\ref{fig12} illustrates how the battery lifespan changes based on the notification time ($T_{\text{notification}}$). As expected, our results show that longer $T_{\text{notification}}$ increases battery life. Specifically, when using 5-minute intervals, the battery remains operational for approximately 2.3 and 3 years for DR8 and DR9, respectively. However, with a 30-minute notification interval, the battery's lifespan exceeds 4.2 years. 

At the time of our experiments, only LR1120 model development kits were commercially available.  As discussed in Section~\ref{LR1120}~\textit{LR1120 development kit} and confirmed by our measurement results in Fig.~\ref{fig6} and Fig.~\ref{fig7}, the time duration and current consumption change across different operational states. The current consumption in the \textit{transmission} (S\_TX) mode may exhibit variations in future LR-FHSS models, particularly if they incorporate different power amplifiers and radio designs. These possible variations can be easily added to our model with slight modifications. LR1121~\cite{LR1121Datasheet} is a brand-new LR-FHSS device, it exhibits the same transmit mode power consumption characteristics as LR1120 (refers to Table 3-6: Transmit Mode Power Consumption of LR1120 in~\cite{LR1120Datasheet} and LR1121 in~\cite{LR1121Datasheet}). We expect our ToA model to remain valid for all the LR-FHSS devices, regardless of potential variations in current consumption. 

\section{Conclusion}
\label{sec:7}
LR-FHSS is emerging as a potential connectivity solution for battery-powered and tiny satellite IoT devices operating in hard-to-reach areas. Frequently replacing a device's battery in remote industrial or fragile natural areas, e.g., offshore wind farms to give just one illustrative example, is challenging and costly. The study of the LR-FHSS time and energy profiles provides important insights for further communication and application operation optimizations; it is also crucial for understanding this technology's feasibility and potential performance for practical applications and use cases.
 
The main contributions of this paper are the LR-FHSS current consumption and the ToA analytical model based on real experimental measurements. First, we conduct extensive measurements for variable transmit power, Date Rates, and message payload size. The results offer valuable insights into the various system states associated with LR-FHSS transmissions, including their respective timing and current consumption characteristics. Importantly, this study identifies additional states, such as \textit{Radio Preparation} and \textit{Radio off}, which were not previously reported in the LR-FHSS-enabled LR1120 development kit user manual.

Notably, our proposed ToA model demonstrates better accuracy than existing state-of-the-art models. Similarly, this work introduces the first analytical current consumption model for LR-FHSS. More than this, our paper discusses the average current consumption and battery lifetime as a function of variable transmit power and the transmission notification time. Our models and results can be useful for further studies investigating the medium access control aspects and battery lifetime of LR-FHSS-enabled devices. They enable us to study the feasibility and performance of the applications (e.g., industrial use cases) based on them. \textcolor{black}{In future, we aim to expand our experimental measurements to examine the current and ToA characteristics of forthcoming LR-FHSS device models.} Also, we consider comparing the current consumption and battery life of LoRa, NB-IoT, and RedCap technology as one research direction.
\section*{Acknowledgments}
This research was supported by the Research Council of Finland (former Academy of Finland) 6G Flagship Programme (Grant Number: 346208). The studies of K. Mikhaylov were supported by the Research Council of Finland (former Academy of Finland) MRAT-SafeDrone (341111) project. We thank the Finnish foundations, including Walter Ahlström, Tauno Tönningin, Riitta ja Jorma J. Takasen and Nokia Foundation for supporting this study.
\bibliographystyle{IEEEtran}
\bibliography{LRFHSSModelsPaper.bib}
\end{document}